\documentclass[aps,prb,10pt,twocolumn, superscriptaddress, longbibliography]{revtex4-1}
\usepackage{amsmath}
\usepackage{amsfonts}
\usepackage{amssymb}
\usepackage{graphicx}
\usepackage{bm}
\usepackage[colorlinks=true,linktocpage=true,breaklinks=true,
allcolors=blue]{hyperref}
\usepackage{xcolor}

\begin{document}

\title{Ultrafast optical generation of antiferromagnetic meron-antimeron pairs with conservation of topological charge}

\author{Sumit Ghosh}
\email{s.ghosh@fz-juelich.de}
\affiliation{Peter Gr{\"u}nberg Institut (PGI-1), Forschungszentrum J{\"u}lich GmbH, 52428 J{\"u}lich, Germany}
\affiliation{Institute of Physics, Johannes Gutenberg-University Mainz, 55128 Mainz, Germany}

\author{Stefan Bl\"ugel}
\affiliation{Peter Gr{\"u}nberg Institut (PGI-1), Forschungszentrum J{\"u}lich GmbH, 52428 J{\"u}lich, Germany}

\author{Yuriy Mokrousov}
\affiliation{Peter Gr{\"u}nberg Institut (PGI-1), Forschungszentrum J{\"u}lich GmbH, 52428 J{\"u}lich, Germany}
\affiliation{Institute of Physics, Johannes Gutenberg-University Mainz, 55128 Mainz, Germany}

\begin{abstract}
 We propose a new mechanism to produce a meron-antimeron pair in a two dimensional antiferromagnet with an ultrafast laser pulse via thermal Schwinger mechanism. Unlike ultrafast skyrmion nucleation, this process conserves total topological charge. We systematically show different stages of the dynamics and define proper topological invariants to characterise the configurations. The emergent structure can retain its topological structure for up to 100\,ps. By introducing a topological structure factor we show that pair formation is robust against any random choice of initial magnetic configuration and can survive against disorder. Our findings demonstrate that the rich world of spin textures, which goes beyond conventional skyrmions, can be reached optically.  
\end{abstract}

\maketitle

\section{Introduction}

In the quest for a fast and efficient mechanism for manipulating magnetic order, ultrafast optical manipulation of magnetism emerged as one of the most promising paradigms. Interaction between an ultrafast laser and magnetic moments takes place at a much faster timescale and consumes far less energy \cite{Kirilyuk2010, Radu2011} compared to electrical switching \cite{Yang2017}. The underlying physical mechanism is, however, still in a mist. From the initial experimental demonstration of ultrafast demagnetisation \cite{Beaurepaire1996} this effect was considered to be of thermal origin conventionally modelled with the phenomenological three temperature model. Zhang and H\"ubner \cite{Zhang2000} proposed a microscopic mechanism mediated by spin-orbit coupling, whereas Koopmans and co-workers proposed an alternative mechanism mediated by electron-phonon coupling \cite{Koopmans2005}. These theoretical frameworks aim to understand the transition from ordered magnetic states to strongly disordered and often nonmagnetic states, which  gives the impression that the governing mechanism behind the ultrafast demagnetisation is of thermal origin \cite{Koopmans2010}. Based on time-dependent density functional theory approach, Dewhurst {\it et al.} recently proposed an alternative mechanism \cite{Dewhurst2018} for transferring angular momentum between different sites,which has been subsequently observed experimentally~\cite{Willems2020}. This mechanism relies on a coherent redistribution of spin angular momentum \cite{Dewhurst2021} and therefore can be exploited to drive a ferromagnet-to-antiferromagnet transition \cite{Golias2021}. 

In this context, an ability to generate any desired magnetic configuration,~i.e.~a spin texture of desired properties, with an ultrafast optical laser, remains a challenge. Recent experimental demonstration of skyrmion nucleation with an ultrafast laser by B\"uttner {\it et al.}~\cite{Buttner2021} has proved the feasibility of optical generation of spin textures. However, the proposed explanation of the effect relies solely on classical magnetisation dynamics which does not take into account electronic interactions. Besides, the working model contains the anti-symmetric Dzyaloshinskii-Moriya interaction which is known to drive the system into a skyrmionic state without any additional excitation~\cite{Fert2013}. The nucleation process can be further reinforced with a laser~\cite{Bostrom2022} in the presence of spin-orbit coupling. Such topological structures are also known to arise during the transition between two different magnetic configurations \cite{Heo2016, Ono2019}, however, their transient nature makes them useless for any practical purpose. An alternative mechanism of laser induced magnetisation dynamics was recently proposed by Ghosh and co-workers~\cite{Ghosh2022}, showing the emergence of a new chiral spin-mixing interaction which can lead to chiral formation even in absence of any intrinsic spin-orbit coupling. The emergent chirality is quite sensitive to the laser parameters which emphasizes the importance of coherent interaction of the laser with the electronic degrees of freedom for chiral state generation. This mechanism leads to a quasi-stable chiral formation which can survive for several picoseconds, and it is distinctly different from that driven by  thermal excitations.

\begin{figure}[t!]
\centering
\includegraphics[width=0.48\textwidth]{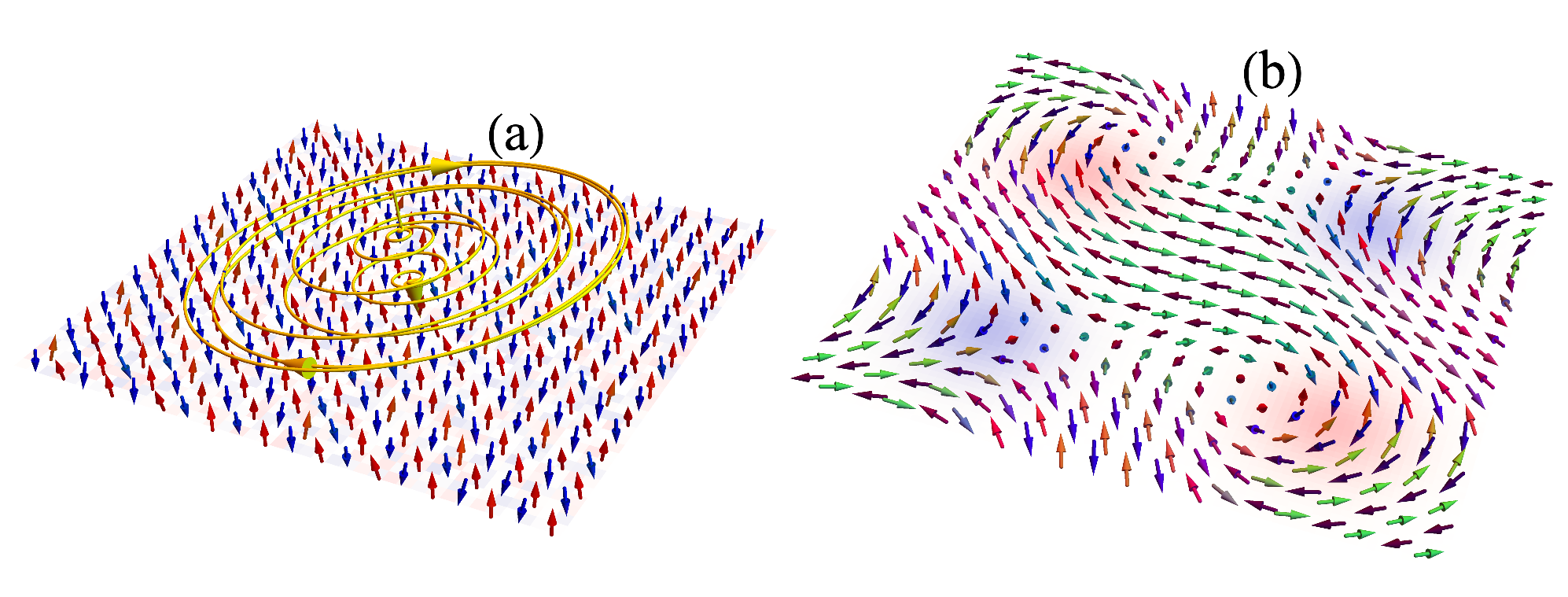}
\caption{Generation of a meron-antimeron pair by an ultrafast laser pulse. (a) Initial antiferromagnetic configuration. Yellow line denotes the circularly polarised laser. (b) Emergence of a quasi-stable meron-antimeron pair after the action of the laser pulse. The blue and red regions show the negative and positive topological charge.}
\label{fig:meron}
\end{figure}

It is well-known that skyrmions are nontrivial magnetic configurations that are equivalent to magnetic monopoles which can be characterised by a skyrmion number \cite{Nagaosa2013}. To understand the dynamics of the skyrmions or similar quasi particles, one can look into the physics of monopoles from the perspective of field theory  \cite{Preskill1984}. Due to their heavy mass~\cite{Saha1936, Goud2017}, creating isolated magnetic monopoles still remains quite challenging. An alternative paradigm would lie in  producing a monopole-antimonopole pair~\cite{Kobayashi2021}, which is also making its appearance in recent experimental observations of magnetic textures. Indeed,  topological spin textures often emerge in pairs consisting of a texture and its anti-texture, keeping the total topological charge to zero. Such pair formation has been observed both at metallic interfaces~\cite{Gao2019} and predicted in two-dimensional magnets~\cite{Lu2020, Augustin2021}. These pairs can be further transmuted to a pair of different but opposite textures (for example meron to skyrmion \cite{Yu2018} conversion by an external magnetic field) without violating the conservation of topological charge.

All these studies so far have been done on ferromagnetic materials which naturally raises an alluring question $-$ is it possible to combine the aforementioned ideas and stimulate such pair production in an antiferromagnet via an optical excitation? Antiferromagnets are favourable materials due to their natural abundance and faster response as compared to ferromagnets. Besides, they are immune to any external magnetic field and also demonstrate rich and topologically nontrivial electronic properties. However, regarding the nontrivial real space textures in antiferromagnets, reported studies are restricted to a narrow domain of skyrmions only  \cite{Zhang2016, *Buhl2017, *Gobel2017}. Here, we demonstrate that it is indeed possible to go beyond skyrmions and nucleate meron-antimeron pairs in an antiferromagnet with an ultrafast laser pulse (Fig.\,\ref{fig:meron}). We present a systematic study of the different stages of nucleation and demonstrate the gradual formation of meron-antimeron pair starting from a collinear antiferromagnetic state. We analyze different physical observables at different stages of the dynamics and define a proper topological invariant to characterise their dynamical evolution. Finally we reveal the robustness of the overall process versus disorder.

\section{Model and Method}

We define a two-dimensional antiferromagnet on a square lattice with a time dependent double-exchange Hamiltonian~\cite{Koshibae2009, Ono2017, Ghosh2022}
\begin{eqnarray}
H(t) &=& -J\sum_{i;\mu,\nu} c^{\dagger}_{i,\mu}(\hat{\bm m}_i \cdot \bm{\sigma})_{\mu\nu} c_{j,\nu} - \sum_{i,j;\mu} h_{ij}(t) c^{\dagger}_{i,\mu}c_{j,\nu},\quad
\label{h1}
\end{eqnarray}
where $c^{\dagger}_{i,\mu},c_{i,\mu}$ are the creation (annihilation) operators of an electron at site $i$ with spin $\mu$. $\hat{\bm{m}}_i$ is a unit vector at site $i$ denoting the direction of the magnetisation and $\bm{\sigma}$ is the vector of Pauli spin matrices. $J$ denotes the coupling between the local magnetic moment and the itinerant moment, which, following Hund's rule is kept positive \cite{Ono2017} and set to 1\,eV. $h_{ij}(t)$ is the time-dependent hopping parameter between site $i$ and site $j$. The action of the laser is defined as a time varying electric field with a Gaussian envelope modelled as a time varying vector potential $\bm{A}(t)$ = $-\frac{\mathcal{E}_0}{\omega}e^{-(t-t_0)^2/2s^2}(\cos(\omega (t-t_0))\hat{\bm x} + \sin(\omega (t-t_0))\hat{\bm y})$, where $\mathcal{E}_0$ is the peak amplitude of laser occurring at time $t$=$t_0$ and $\omega$ is the angular frequency. $s$ denotes the broadening of the pulse which we kept at 5\,fs. In this work we keep $\mathcal{E}_0$=$0.1$\,eV$\cdot a_0^{-1}$ ($a_0$ being the lattice constant), and $\omega$=$2.02\,|J|/\hbar$=$3.07 \times 10^{15}$\,Hz. The frequency is chosen such that the laser connects the unoccupied states with the occupied states which are separated by an energy gap of 2$|J|$. The action of the laser is incorporated in the Hamiltonian (Eq.\ref{h1}) via Pierel's substitution $h_{ij}(t)$=$h e^{i(e/\hbar)\bm{A}(t) \cdot \bm{r}_{ij}}$, where $r_{ij}$ is the vector connecting site $i$ to site $j$. The static value of the hopping parameter is kept at $h$=$0.4$\,eV. The qualitative behaviour of the system remains consistent within a moderate range of these parameters \cite{Ghosh2022}. For our study we consider a super-cell of dimension $20 \times 20$ with periodic boundary condition.

\begin{figure}[t!]
\centering
\includegraphics[width=0.48\textwidth]{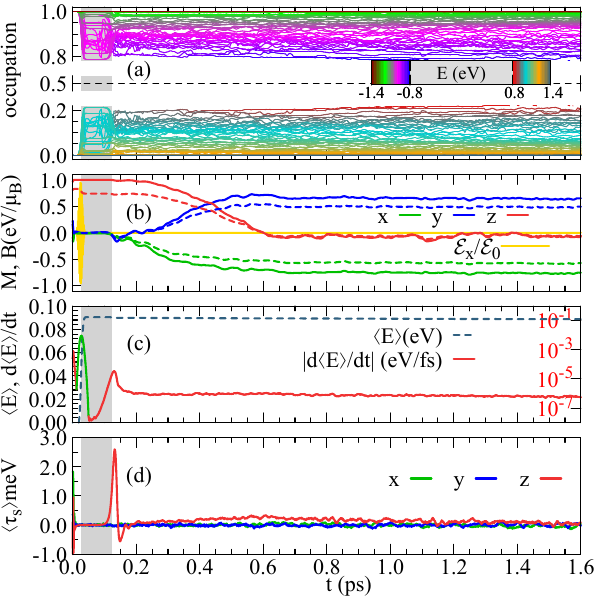}
\caption{Time evolution of different physical observables. (a) Change of occupation for different states with their colour denoting corresponding eigenvalues. Here we show every 5th state. Gray shaded region shows the time delay between the incident of laser and onset of magnetisation dynamics. (b) Change of magnetisation vector (solid) and effective field (dashed) components of the magnetic moment at  $x$=1,$y$=10. Yellow line denotes the $x$ component of the laser electric field. (c) Change of average total energy (gray dashed) and rate of change of energy (red) where the green portion shows when the system absorbs energy. (d) Evolution of different components of average torque.}
\label{fig:mt}
\end{figure}

The ground state is constructed by filling half of the lowest eigenvalue states which in our case are all the states with negative eigenvalues. For our choice of parameters, this corresponds to an antiferromagnetic alignment \cite{Koshibae2009, Ghosh2022}. We simulate the thermal equilibrium at finite temperature by adding a small random fluctuation (0.1$\pi$) to the polar angles while the azimuthal angles are chosen randomly between 0 and 2$\pi$. This randomness is crucial to initiate the magnetisation dynamics by generating initial spin-mixing interaction \cite{Ghosh2022}. Note that our initial configuration does not posses any spin-orbit coupling or any specific non-collinear order and therefore constitutes a trivial antiferromagnetic insulator. At any instance of time $t$ any quantum state $|\psi_l\rangle$ can be expressed as the linear combination of instantaneous eigenstates $|n(t)\rangle$ of the Hamiltonian which is evolved within Schr{\"o}dinger picture \cite{Koshibae2009, Ghosh2022}. This allows us to evaluate the instantaneous effective field for any site $i$ as $\frac{1}{\mu_B}\sum \langle \psi_l |- \nabla_{\bm{m}_i} H| \psi_l \rangle $, $\mu_B$ being the Bohr magneton. This effective field is used in a set of Landau-Lifsitz-Gilbert equations \cite{Ono2017, Petrovic2018,  Ghosh2022} 
\begin{eqnarray}
\frac{d\bm{m}_i}{dt} = -\gamma (\bm{m}_i \times \bm{B}_i) - \lambda \bm{m}_i \times (\bm{m}_i \times \bm{B}_i),
\end{eqnarray}
where $\gamma = \frac{g_e \mu_B}{\hbar \mu_s}\frac{1}{1+\alpha^2}$, $\lambda=\frac{g_e \mu_B}{\hbar \mu_s}\frac{\alpha}{1+\alpha^2}$, $g_e$ is the gyromagnetic ratio, $\mu_s$ is the onsite magnetic moment which we keep at 1\,$\mu_B$ and $\alpha$ is the dimensionless damping coefficient which we keep fixed at 0.2. $\bm{m}_i \times \bm{B}_i$ constitutes the local torque. One can add an additional stochastic torque \cite{Garcia1998} to simulate thermal fluctuation. For simplicity we neglect such a thermal effect since the topological nature of the outcome remains robust at low temperature. 

\section{Result and Discussion}

After the action of the ultrafast laser, one can observe three different phases of dynamics originated from complex interactions between electronic and magnetic degrees of freedom (Fig.\,\ref{fig:mt}). The peak amplitude of the laser occurs at $t=25$\,fs (Fig.\,\ref{fig:mt}b), marked by the left edge of the gray area in Fig.~\ref{fig:mt}. This triggers a redistribution of occupation of quantum states (Fig.\,\ref{fig:mt}a) which in turns start building the torque. An initial randomness is imperative to start this process which also determines how fast the system can respond to the pulse  \cite{Ghosh2022}. In our present setup  after a latency period of approximately 100\,fs (right edge of gray region in Fig.\,\ref{fig:mt}) the magnetic moments start re-organising themselves causing the fast relaxation phase \cite{Ghosh2022} that lasts for approximately 0.5\,ps (Fig.\,\ref{fig:mt}b). The system then enters into the slow relaxation phase which eventually leads to the final texture shown in Fig.\,\ref{fig:meron}. Note that the energy absorption takes place only during the short duration defined by the Gaussian envelope of the pulse (green segment in Fig.\,\ref{fig:mt}c). The steady state can be characterised by evaluating the average staggered torque \cite{Ghosh2019} defined as $\langle \tau_s \rangle = 1/N \sum_{ij} \nu_{ij} ( \bm{m}_{ij} \times \bm{B}_{ij})$, where the index $i,j$ correspond to the position, $\nu_{ij}=\pm 1$ is the sublattice index, and $N$ is the total number of sites. Note that the optical excitation predominantly produces the out of plane component of the torque (Fig.\,\ref{fig:mt}d) leading to the final in-plane texture. Assuming that the average energy gain per site is $\sim$\,0.1\,eV with the dissipation rate being $~10^{-6}$\,eV\,fs$^{-1}$, it would take the system about 100\,ps to dissipate the excess energy and regain its initial configuration. It is worth mentioning that due to the complex intertwined electronic and magnetisation dynamics, the energy dissipation of the system is not solely determined by the phenomenological damping parameter only and additional dissipation channels can emerge during different stages of the dynamics. This can be verified from the energy dissipation rate of the system at steady state which is found to decreases with the increase of damping parameter for such laser excited system~\cite{Ghosh2022}.  

\begin{figure}[t!]
\centering
\includegraphics[width=0.42\textwidth]{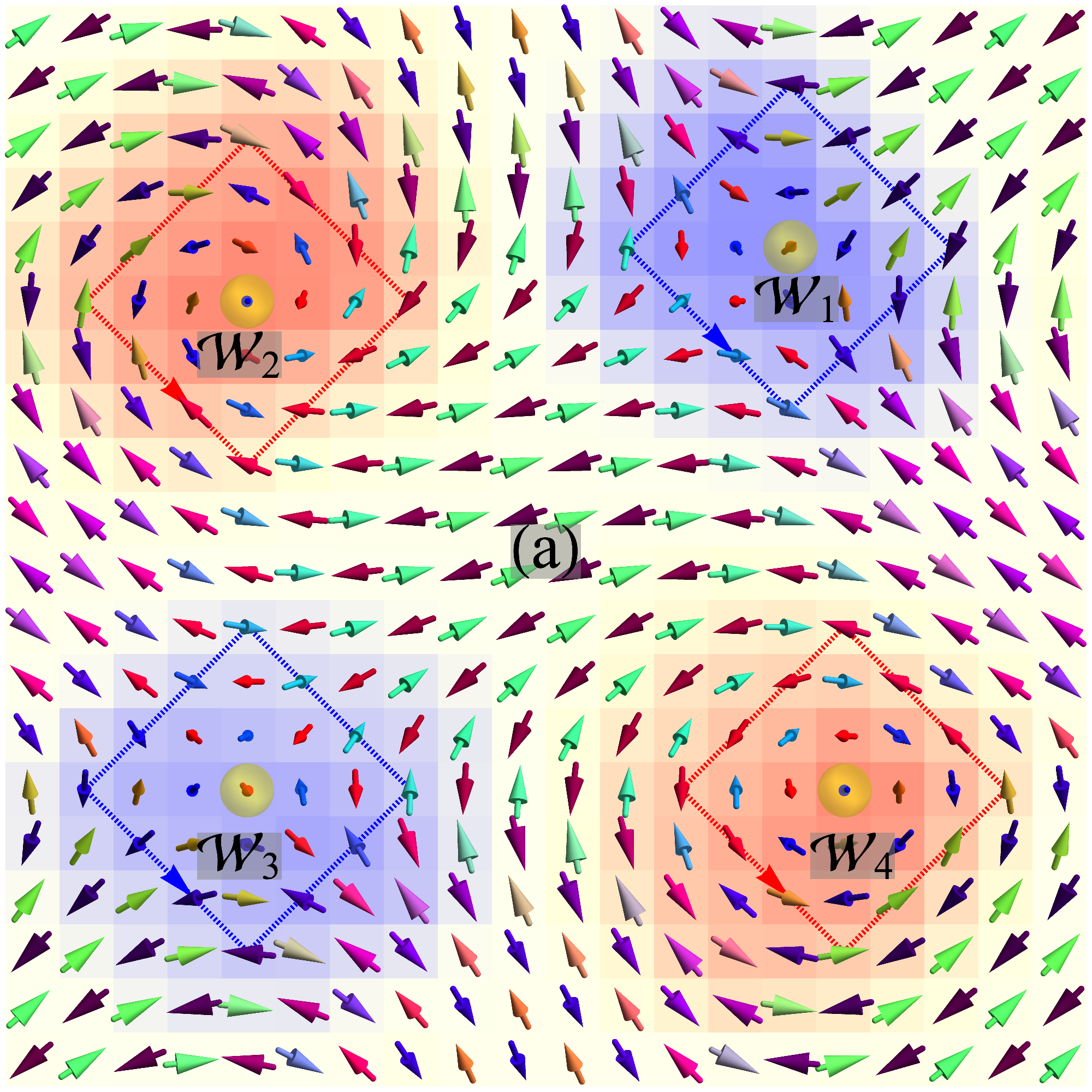}\vspace{0.3cm}
\includegraphics[width=0.45\textwidth]{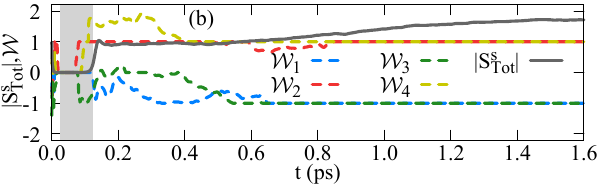}
\caption{Skyrmionic charge and winding number for antiferromagnetic meron-antimeron pairs. (a) Distribution of skyrmionic charge is shown with red and blue colour denoting positive and negative values, with winding centres $\mathcal{W}_{1,2,3,4}$ marked with yellow spheres. Solid blue and red lines show the contours along which winding number is calculated with red and blue colours denoting a winding number $\pm$1. (b) Time evolution of total absolute skyrmion charge density and winding number.}
\label{fig:skw}
\end{figure}

\subsection{Dynamical evolution of topological charge}

From Fig.\,\ref{fig:meron} one can see that the final configuration contains a pair of antiferromagnetic meron and anti-meron. This is reminiscent of Schwinger mechanism which produces a particle and anti-particle. Similar mechanism can also generate monopole-antimonopole pairs \cite{Affleck1982, *Affleck1982a},  that can survive even in presence of a thermal bath \cite{Gould2019}. Such pair nucleation for topological defects was also predicted in the context of bubble collision \cite{Digal1996, *Copeland1996, *Digal1997} following Kibble mechanism \cite{Kibble1976}. To characterise the pairs we choose their skyrmionic charge.  On a ferromagnetic background, each meron can be characterised by a skyrmion number $\pm \frac{1}{2}$, resulting in a total zero skyrmion number for the whole configuration. For an antiferromagnet, one can define a skyrmion number for each sublattice~\cite{Ono2019}, however it does not take the inter-sublattice interaction properly into account. To avoid this, we define a staggered skyrmionic charge (Fig.\,\ref{fig:skw})
\begin{eqnarray}
S^s_{ij} = \nu_{ij} [\bm{m}_{ij} \cdot (\partial_x \bm{m}_{ij} \times \partial_y \bm{m}_{ij})],
\end{eqnarray}
where $\nu_{ij}$=$\pm$1 for the opposite sublattices and $\partial_x \bm{m}_{ij}=(\bm{m}_{i+1,j}-\bm{m}_{ij})$. Since the total skyrmionic charge of the complete cell is zero, we use an absolute skyrmion density $S^s_{Tot}=\sum |S^s_{ij}|$ to characterise the configuration. For a well-isolated meron-antimeron pair this number should be 1. However due to the finite size there is always some overlap between the different regions causing a deviation of the total absolute skyrmion charge from 2 for the whole super-cell (Fig.\,\ref{fig:skw}b). For a proper topological characterisation in this case, we evaluate the winding number \cite{Rybakov2021Th, *Kuchkin2020a} for all four meron/antimeron centres defined as 
\begin{eqnarray}
\mathcal{W} = \frac{\mathcal{P}_s}{2 \pi} \oint_\mathcal{C}  d\bm{l} \cdot \bm{\nabla}\phi
\end{eqnarray}
where $\mathcal{P}_s$ is the sub-lattice polarisation, given by sgn$(m_z) \cdot \nu$ with $m_z$ being the out of plane component of the magnetisation at the centre and $\nu=\pm 1$ for the sub lattices. $\phi$ is the azimuthal angle and the contour $\mathcal{C}$ encircles the winding centre in anticlockwise direction passing through only one type of sublattice. We choose four centres (Fig.\,\ref{fig:skw}a), each specifying a meron/antimeron and calculate the corresponding winding number over the entire evolution period (Fig.\,\ref{fig:skw}). The winding numbers, although initially showing some fluctuating non-integer values originating from the lack of well defined winding centres during the fast and slow relaxation process, eventually attain a steady magnitude of $\pm1$ for $t\gtrsim 0.6$\,ps. Compared to that, the total absolute staggered skyrmion charge saturates slowly towards a magnitude of 2, indicating the presence of four half instantons. It is worth mentioning that the winding number remains unchanged under small changes in the winding center as long as path encircles it. Therefore, the result remains consistent with the skyrmionic charge distribution even when the winding centre differs from the  centre of topological charge.

\begin{figure}[t!]
\centering
\includegraphics[width=0.48\textwidth]{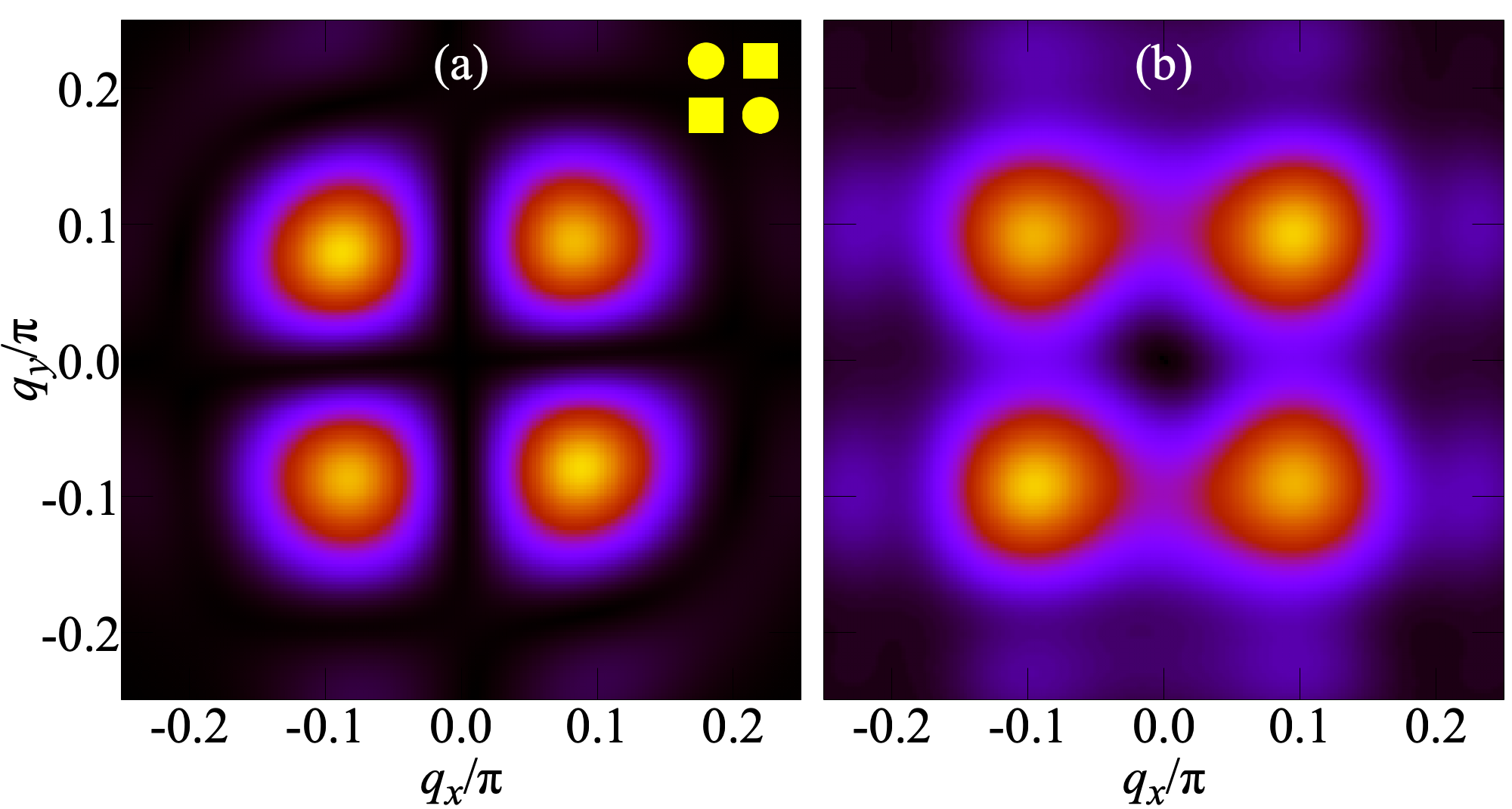}
\includegraphics[width=0.48\textwidth]{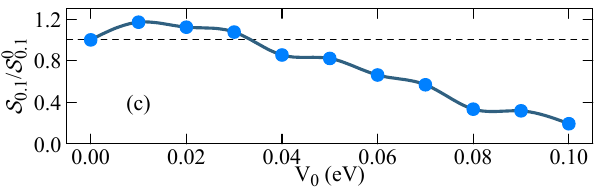}
\caption{Topological structure factor and its stability. (a) Topological structure factor of the configuration shown  in Fig.\,\ref{fig:skw}a. Inset shows the corresponding real space distribution of the topological charge. (b) Average topological structure factor averaged over 70 different configurations. (c) Average topological structure factor for different strength of scalar impurity averaged over 32 different configurations and scaled with its average value at $V_0=0$. For each configuration we consider the mean value at $q=0.1(\pm \hat{x} \pm \hat{y})\pi$.}
\label{fig:sf}
\end{figure}

\subsection{Topological structure factor}
 
The evolution of absolute staggered skyrmion charge and winding number ensures the emergence of topological objects with opposite charges coming in pairs, however it does not yet specify the distribution of the topological charge. To determine that we define a topological structure factor
\begin{eqnarray}
\mathcal{S}_{\bm{q}} = \frac{1}{\mathcal{N}}\sum_{ij} S^s_{\bm{r}_i} S^s_{\bm{r}_j} e^{i \bm{q} \cdot (\bm{r}_i - \bm{r}_j)}
\end{eqnarray} 
where $\mathcal{N}=\sum_{i}(S^s_{\bm{r}_i})^2$ is the normalisation constant. One can readily see from Fig.\,\ref{fig:sf}a that the topological structure factor is reminiscent of a checker-board pattern (Fig.\,\ref{fig:sf}a inset) which is consistent with the skyrmionic charge distribution (Fig.\,\ref{fig:skw}). However, it is also possible that the laser excited system moves into a spin-spiral configuration or a combination of random clusters with no global long range order. To verify that this is not the case, we consider 70 different configurations with the same normalisation constant and evaluate the average distribution of the topological structure factor (Fig.\,\ref{fig:sf}).  The obtained maxima in the distribution occur for $\bm{q}= 0.1(\pm \hat{x} \pm \hat{y})\pi$ which establishes the emergence of a meron-antimeron pair in our $20 \times 20$ super cell. 

We finally study the robustness of these emergent configuration against disorder. For that, we introduce random impurities modelled by on-site energy deviations ranging from $-V_0$ to $V_0$, and measure the topological structure factor at points $\bm{q}=0.1(\hat{x} \pm \hat{y})\pi$ (Fig.\ref{fig:sf}). One can readily see that the structure manages to retain its basic features for fairly strong values of the impurity potential, and then gradually collapses into a featureless distribution. Note that a small random impurity can enhance the formation as denoted by the initial rise in average topological structure factor which is consistent with the behaviour against small random fluctuation of magnetic moment. This establishes the robustness of the pair formation process which would be helpful for its physical realisation.

\section{Conclusions}

In this work we present a new paradigm for generating non-trivial magnetic textures in antiferromagnets. We show that when excited by an ultrafast laser a two-dimensional antiferromagnet can host meron-antimeron pairs via thermal Schwinger mechanism. The topological texture can survive for up to 100\,ps which makes it possible to observe experimentally. The process is independent of initial helicity of the laser and responsive within a moderate range of frequency and peak amplitude of the laser. We define a staggered skyrmionic charge,  winding number and topological structure factor to demonstrate the formation of topological charge and show that on average the system is more likely to form a pair of meron and anti-meron in a checker board pattern. This formation is fairly robust against scalar impurity which makes it easier to observe experimentally. Since the total topological charge of a texture / anti-texture pairs is always zero, pairs of different genera can be connected smoothly allowing more flexibility in exploring topological objects of higher genus. Our results thus open a new route to generate out-of-equilibrium topological features which remain inaccessible in their ground state configuration. 
This takes us beyond the generation of conventional nontrivial structures such as skyrmions and domain walls, and paves the way to generating more exotic higher-order topological objects such as chiral bobbers~\cite{Rybakov2015} or hopfions \cite{Rybakov2019}.

\section{Acknowledgments}

We thank Nikolai Kiselev, Stefan Eisebitt, Frank Freimuth and Olena Gomonay for discussions.  
We acknowledge financial support from Leibniz Collaborative Excellence project OptiSPIN $-$ Optical Control of Nanoscale Spin Textures.
We  acknowledge  funding  under SPP 2137 ``Skyrmionics'' of the DFG and financial support from the European Research Council (ERC) under the European Union's Horizon 2020 research and innovation program (Grant No. 856538, project ``3D MAGiC''). The work was also supported also by the Deutsche Forschungsgemeinschaft (DFG, German Research Foundation) $-$ TRR 173/2 $-$ 268565370 (project A11), TRR 288 $-$ 422213477 (project B06).  We  also gratefully acknowledge the J\"ulich Supercomputing Centre and RWTH Aachen University for providing computational resources.

%
%

\bibliographystyle{apsrev4-1}
\bibliography{ref}
\end{document}